# On the Brittle-to-Ductile Transition of the As-cast TiVNbTa Refractory High-entropy Alloy

Robert J. Scales[1], David E.J. Armstrong[1], Angus J. Wilkinson[1], Bo-Shiuan Li[1]

[1]*Department of Materials, University of Oxford, Oxford OX1 3PH, UK*

*Corresponding Author: Robert James Scales,*
*Email: robert.scales@mansfield.ox.ac.uk,*
*Full postal address: Department of Materials, University of Oxford, Oxford OX1 3PH, UK*

**Abstract:**
The fracture properties of as-cast TiVNbTa, a refractory high-entropy alloy (**RHEA**), were investigated using four-point bending tests from -139°C to 20°C under a strain-rate of $10^{-3}$ $s^{-1}$. From those tests and fractography, the conditional fracture toughness values and the brittle-to-ductile transition temperature were obtained. The brittle-to-ductile transition temperature was between -47°C to -27°C, which gives an estimated activation energy value of 0.52±0.09 eV for the alloy. This study provides a preliminary understanding of the nature of dislocation motion in a relatively ductile **RHEA**.

**Keywords:**
1. High-entropy alloys
2. Refractory metals
3. Fracture toughness
4. Brittle-to-ductile transition
5. Bending test

**Main Body**

High-entropy alloys (**HEAs**) are a class of crystalline metallic material which does not have an unambiguously identifiable base element and is highly alloyed with multiple elements. Some HEAs have demonstrated the potential of exhibiting superior mechanical properties compared to conventional alloys [1–3]. This has recently energised the metallurgy community by providing a vast elemental palette for alloy design. HEAs have also generated significant interest in the energy and aerospace sectors, as there is always a strong demand for more efficient power-generating systems, which can be achieved by making the engineering components operate at higher temperatures for longer durations. Therefore, new alloy design strategies for HEAs have been proposed with the ultimate goal of surpassing existing steels and Ni-based superalloys for nuclear and aerospace applications [4–6]. This new class of HEA consists of mainly refractory elements and exhibit a body-centred cubic (**BCC**) structure, and are termed as refractory high-entropy alloys (**RHEAs**) [7,8].

Achieving high-temperature strength for RHEAs has been relatively successful, for example, the yield strength for NbMoTaW and VNbMoTaW at 1600°C are 405 MPa and 477 MPa respectively, which are values superior to current superalloys [8]. However, the main issue for RHEAs is their lack of room temperature ductility [9]. This is often attributed to their severely distorted lattice (from alloying many elements of similar size) but can also result from the frequent precipitation of brittle intermetallic phases; both of which act as barriers for dislocation motion. The equiatomic TiVNbTa is one of the few single-phase BCC RHEAs which exhibits both excellent high-temperature strength and compressive ductility [10,11]. Its yield strength at room temperature is 1273 MPa, which drops to 688 MPa at 900°C, and it has a compressive strain to failure of over 30% [11]. However, nearly all of the reported mechanical data were from compression tests, where fracture and catastrophic failure is suppressed. Tensile or bend tests are rarely conducted but are far more critical to engineering applications as they allow for fracture properties to be determined in a tensile stress state. So far only a few tensile tests have been conducted for RHEAs and were only done at room temperature [12]; the lack of non-ambient tensile mechanical properties is the main limiting factor for RHEAs to be used in engineering applications.

The deformation behaviour of BCC metals is strongly dependent on both temperature and strain-rate because dislocations are a carrier of plasticity, and dislocation motion is both thermally-activated and rate-sensitive [13]. At a given strain-rate, a brittle-to-ductile transition temperature (**BDTT**) is often used to describe the critical temperature for the transition in deformation behaviour. However, in most engineering alloys, pre-existing dislocations and complicated microstructures cause the transition to occur gradually rather than the sharp transitions seen in pristine single crystals [14]. In addition, interstitial impurities and alloying elements are known to have strong influence on the BDTT. However, it is yet unclear how the complicated alloying elements within the RHEA solid-solution can influence the dislocation motions and the BDTT. Therefore, it is crucial to perform fracture tests over a range of temperature to measure the BDTT of the newly-developed TiVNbTa to prevent catastrophic failure during operation.

The TiVNbTa RHEA was manufactured using an Arc200 arc furnace (Arcast Inc., USA). High-purity (>99.95%) raw elements in lump form were used and were obtained from Goodfellow (Goodfellow Cambridge Ltd., UK). The cast ingot was flipped and remelted multiple times to achieve chemical homogeneity before being tilt-cast into a cylindrical water-cooled copper mould. The composition of the as-cast TiVNbTa RHEA is given in **Table 1**, with its nominal composition being $Ti_{25}V_{25}Nb_{25}Ta_{25}$.

| Element | Ti | V | Nb | Ta | C | N | O |
|---|---|---|---|---|---|---|---|
| at.% | 23.592 | 24.439 | 25.041 | 26.797 | 0.0276 | 0.0073 | 0.0963 |

**Table 1** – The composition of the as-casted TiVNbTa RHEA investigated (in atomic percent), obtained via x-ray fluorescence by AMG Superalloys Ltd.

The as-cast TiVNbTa rod was EDM-machined (electrical discharge machining) into matchstick samples (~1.2 x 1.2 x 11.0 mm). These were then ground using SiC grit papers until all surface flaws were removed and exhibited a smooth finish. Notching of the matchsticks was also done through EDM, using a single-edged tungsten razorblade to generate a blunt notch with sharp thermally-induced cracks. The notch depths were chosen to achieve a depth-to-thickness ratio around 0.1 so that the total lengths of the thermally-induced cracks were consistent throughout samples. The sample geometries are described in **Table 2**, which shows the matchsticks had near square cross-sections, and their notch-to-depth ratios were consistent.

| Sample № | Testing Temperature (°C) | h (mm) | w (mm) | c (mm) | h/w | c/h |
|---|---|---|---|---|---|---|
| A | -139 | 1.208 | 1.186 | 0.127 | 1.02 | 0.10 |
| B | -117 | 1.197 | 1.238 | 0.129 | 0.97 | 0.11 |
| C | -70 | 1.190 | 1.303 | 0.100 | 0.91 | 0.08 |
| F | -60 | 1.179 | 1.191 | 0.130 | 0.99 | 0.11 |
| E | -47 | 1.199 | 1.137 | 0.135 | 1.05 | 0.11 |
| F | -27 | 1.188 | 1.192 | 0.136 | 1.00 | 0.11 |
| G | 19 | 1.114 | 1.147 | 0.136 | 0.97 | 0.12 |
| H | 20 | 1.190 | 1.193 | 0.143 | 1.00 | 0.12 |
| *Average* | | | | | *0.99±0.04* | *0.11±0.01* |

**Table 2** – Matchstick sample geometries and the testing temperatures. The stated uncertainty is the population standard deviation.

X-ray diffraction (**XRD**), scanning electron microscopy (**SEM**), energy dispersive x-ray spectroscopy (**EDX**), and electron-backscattered diffraction (**EBSD**) were carried out for microstructural analysis of the as-cast TiVNbTa. The XRD was done on a MiniFlex 600 (Rigaku, Japan) x-ray diffractometer, using the following parameters: x-rays of wavelength ~1.54 Å (i.e. Cu-Kα1 [15]); a step size of 0.005°; a dwell time of 10 seconds/°, and a scan range of 20°-120°.

SEM, EBSD, and EDX were done using a Zeiss Merlin FEG-SEM (Zeiss, Germany). Four-point bending tests (**4PB**) were carried out in a variable temperature chamber (-150°C to 500°C) integrated within a Shimadzu AGS-X loading rig (Shimadzu UK Ltd., UK). The chamber was connected to a liquid nitrogen dewar for cryogenic testing. The dimensions of the matchsticks were measured before testing. After the target temperature was achieved, a pre-load of 10 N was applied to remove any slack from the load system, then followed by a constant stroke-rate of 15 μm/s until failure or a maximum displacement of 1 mm was achieved.

As shown in **Fig. 1**, the as-cast TiVNbTa exhibits a single-phase BCC crystal structure (a ≈ 3.237 Å, found by manually fitting XRD data with a β-Ti pattern [16]), coarse equiaxed grains, and noticeable dendritic structure, which concurs with observations of the same alloy in literature [4,10]. EDX analysis also revealed slight chemical inhomogeneity, caused by segregation during solidification. All of the bend test samples were cut from the centre of the rod; hence no significant grain size variation was observed.

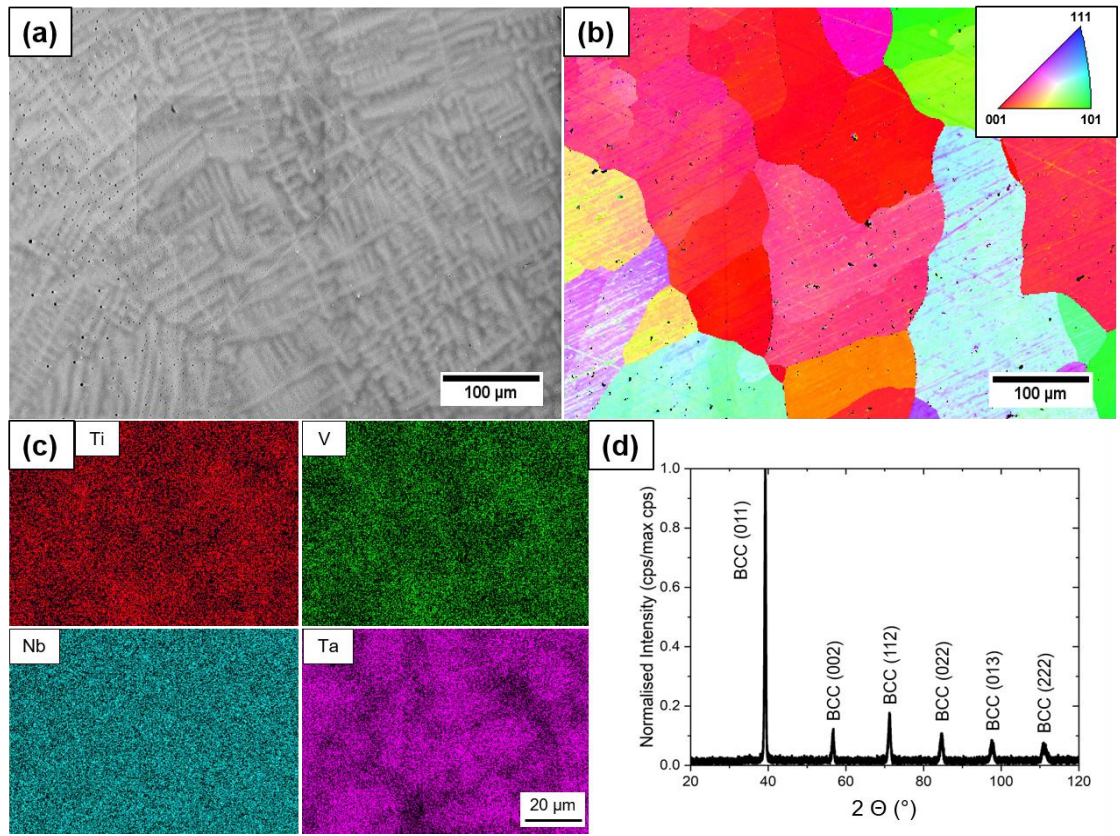

**Fig. 1** – **(a)** Secondary electron micrograph of the as-cast TiVNbTa showing the dendritic microstructure, **(b)** EBSD IPF-Z map showing the large equiaxed grains, **(c)** EDX maps of the constituent elements showing slight micro-segregation, **(d)** XRD pattern, with all the primary peaks indexed as BCC.

4PB tests of EDM-notched as-cast TiVNbTa matchsticks were carried out from -139°C up to 20°C, using a constant stroke displacement rate of 15 μm/s so that the outer surface of the central span achieved a strain-rate of $10^{-3}$ s$^{-1}$. Representative load-displacement curves from the brittle, semi-brittle, and ductile regimes are shown in **Fig. 2(a)**. The brittle matchsticks (-139°C) exhibited tiny, discrete load drops, which indicates limited crack growth and plastic deformation, which was then followed by sudden catastrophic failure. The semi-brittle matchsticks (-70°C, -47°C) exhibited continuous non-linear deformation, indicating a combination of crack growth and plastic deformation, followed by a sudden load drop and then extensive stable crack growth (manifested as continuous load drops). The ductile matchsticks (-27°C, 20°C) exhibited full plastic behaviour out to the 1mm displacement limit without any crack growth.

The conditional fracture toughness values ($K_Q$) were calculated using linear-elastic (**LEFM**) or elastic-plastic fracture mechanical (**EPFM**) analysis shown in **Eqs. 1-4**, depending on whether the failure was brittle or semi-brittle, respectively (as described in the ASTM standards [17,18]). Despite some non-linear deformation, LEFM was used on the brittle matchsticks due to their catastrophic failure. Here, the term "conditional fracture toughness" was used because the semi-brittle matchsticks do not fulfil the specimen size requirements given in ASTM standards [17,18]. The $K_Q$ vs temperature plot is given in **Fig. 2(b)**, showing the soft brittle-to-ductile transition. $K_Q$ was consistent at low temperatures with a value of 43±1 MPa·m$^{0.5}$, which increased to ≥65 MPa·m$^{0.5}$ in the semi-brittle range. At temperatures above -40°C, all matchsticks yielded instead of fractured; therefore, the yielding stress was recorded. This soft transition is typical of semi-brittle BCC alloys, and the BDTT range in the as-cast TiVNbTa was found to be within -47°C to -27°C.

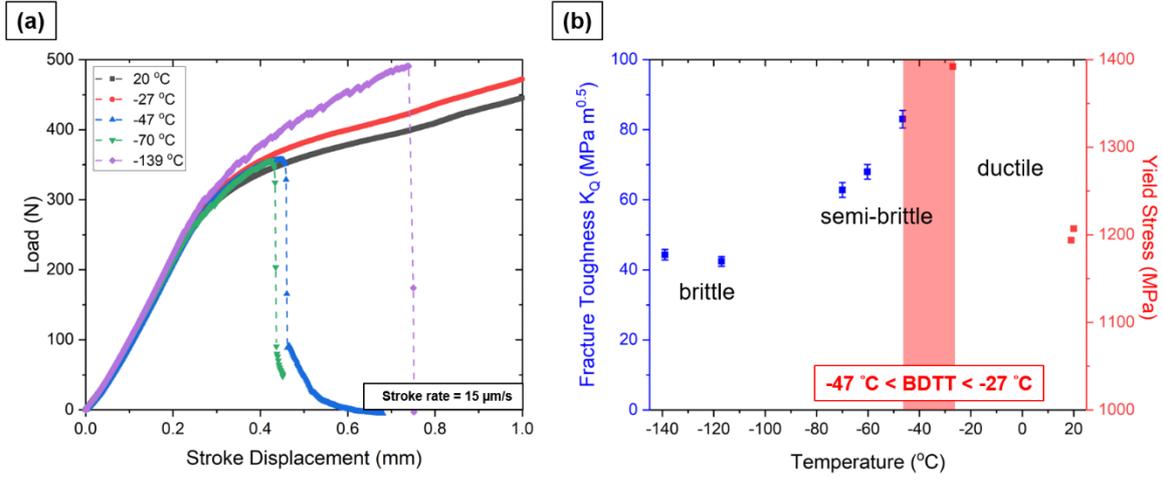

**Fig. 2** – **(a)** Representative load-displacement of the 4PB tests of the as-cast TiVNbTa matchsticks at three temperature regimes: the brittle samples (-137°C, -117°C) exhibited small-scale plasticity and crack growth followed by catastrophic failure; the semi-brittle samples (-70°C, -47°C) exhibited combination of plasticity and stable crack growth; and the ductile samples (-27°C, 20°C) exhibited complete plastic behaviour. **(b)** conditional fracture toughness ($K_Q$) vs temperature plot, showing a soft brittle-to-ductile transition and a BDTT range in between -47°C and -27°C (for the ductile samples, the yield stress is labelled instead of fracture toughness)

$$K_Q = \begin{cases} \sigma\sqrt{\pi c} \cdot \left(1.12 - 1.39\left(\frac{c}{h}\right) + 7.32\left(\frac{c}{h}\right)^2 - 13.1\left(\frac{c}{h}\right)^3 + 14.0\left(\frac{c}{h}\right)^4\right), & LEFM\ (Eq.\ 1) \\ \sqrt{K_{Q,LEFM}^2 + \left(\frac{\eta A_{pl}}{w(h-c)} \cdot \frac{E}{1-\nu^2}\right)}, & EPFM\ (Eq.\ 2) \end{cases}$$

$$\sigma = \frac{3(L_o - L_i)}{2wh^2} \cdot P \quad (Eq.\ 3)$$

$$\eta = \begin{cases} 0.32 + 12\left(\frac{c}{h}\right) - 49.5\left(\frac{c}{h}\right)^2 + 99.8\left(\frac{c}{h}\right)^3, & \frac{c}{h} < 0.282 \\ 2, & \frac{c}{h} > 0.282 \end{cases} \quad (Eq.\ 4)$$

**Equations 1-4** – h = matchstick height, w = matchstick width, $L_o$ = outer span = 10 mm, $L_i$ = inner span = 5 mm, P = applied load, δ = stroke displacement, σ = stress, c = notch depth, $A_{pl}$ = plastic area under the load-displacement curve, E = Young's modulus = 132 GPa (obtained via nanoindentation), and ν = Poisson's ratio = 0.3. **Equations 1-3** are from [17,18] and **Equation 4** is from [19].

The fracture surfaces of the matchsticks tested at each temperature regime were examined via SEM and are shown in **Fig. 3**. The fracture surface of the -139°C matchstick revealed almost entirely cleavage-like features, which is indicative of its brittle failure mode and validates the usage of LEFM. The amount of ductile fracture features (e.g. micro-voids and tearing ridges) on the fracture surfaces increases with increasing temperature, with the -36°C matchstick even showing large areas of both features. The SEM-based fracture surface examination provides clear evidence that the as-cast TiVNbTa undergoes a soft brittle-to-ductile transition as temperature increases (increasing amount of ductile fracture features), which concurs with the BDTT range described in **Fig. 2**.

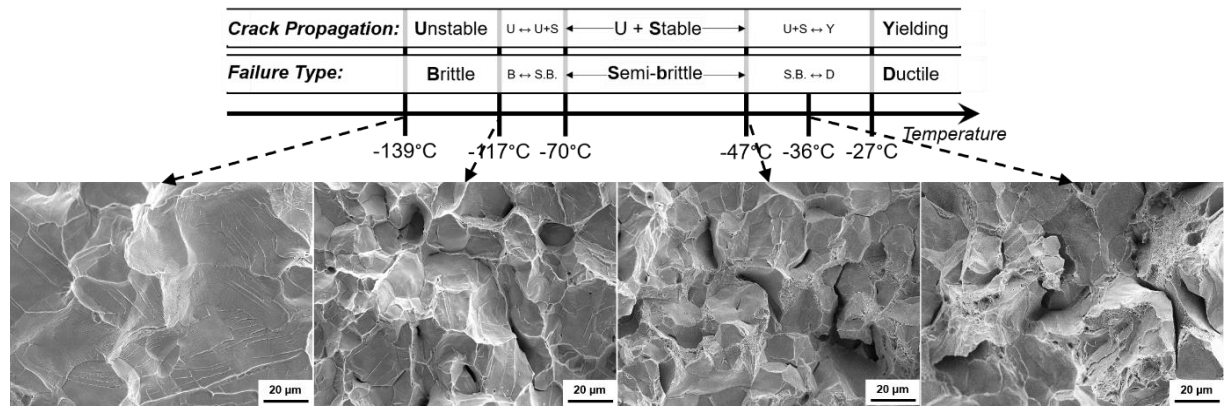

**Fig. 3** – Deformation map and fracture surfaces of the as-cast TiVNbTa matchsticks across different temperature regimes.

**Fig. 4** shows the empirical relationship of activation energy for dislocation glide ($E_{BDT}$) vs BDTT (indicated as $T_{BDT}$ in the graph), as described by Giannattasio et al. [20]. Data from various BCC refractory metals [21–26] have been plotted, and the dotted trend line represents the linear trend described by **Eq. 5** [20]. Therefore, the $E_{BDT}$ for the as-cast TiVNbTa can be estimated, assuming that it follows the same trend, hence producing an $E_{BDT}$ value around 0.522±0.095 eV (when using the BDTT range provided in **Fig. 2**).

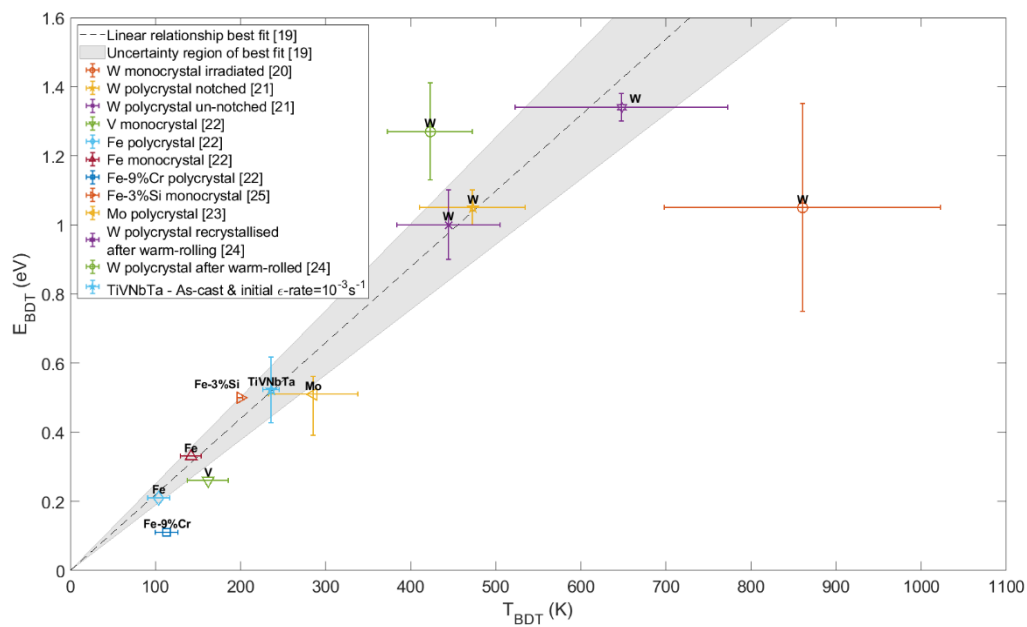

**Fig. 4** – Activation energy ($E_{BDT}$) vs. BDTT ($T_{BDT}$) for various refractory metals. **Eq. 5** ($E_{BDT}/k_B T_{BDT} = 25.5 \pm 3.6$) is plotted as the dashed line and the grey region is the uncertainty [20]. N.B. the y-coordinate of TiVNbTa is calculated using $E_{BDT}/k_B T_{BDT} = 25.5$, with the error bars being the extremes produced using **Eq. 5** and the BDTT uncertainty.

$$\frac{E_A[eV]}{T_{BDT}[K]} = (25.5 \pm 3.6)\frac{k_B}{e} \qquad (Eq. 5)$$

**Equation 5** – $k_B$ = Boltzmann's constant ($1.3806\times10^{-23}$ J/K), and $e$ = elementary charge ($1.6022\times10^{-19}$ C). Equation from Giannattasio et al. [20].

The as-cast TiVNbTa exhibits a good tensile ductility (>5%; maximum strain at the bottom surface of the matchstick) at room temperature despite its high yield strength (~1200 MPa), combined with a relatively low BDTT comparable to BCC metals with engineering application. The lack of tests at other strain-rates prevented the $\dot{\varepsilon}$ vs $1/T_{BDT}$ method to be used to determine a more accurate $E_{BDT}$ [20]. However, an estimated value of around 0.522±0.095 eV can be determined using the empirical relationship given in **Eq. 5**. The BDTT of the as-cast TiVNbTa is similar to that of molybdenum, which was measured experimentally using a similar 4PB setup and strain-rate [24]. The relationship described in **Eq. 5** is sufficiently accurate for most pristine BCC metals but starts to deviate for microstructurally-complicated materials, e.g. warm-rolled and the irradiated tungsten. It is then expected that the highly-concentrated solid-solution in RHEAs and the chemical inhomogeneity in the as-cast state are likely to affect the dislocation motions and the BDTT [27]. Compared to other single-phase BCC RHEAs (NbMoTaW 1.6 MPa·m$^{0.5}$ [28], FeCoNiCrAl$_3$ 7.6 MPa·m$^{0.5}$ [29], and Al$_{18}$Cr$_{21}$Fe$_{20}$Co$_{20}$Ni$_{21}$ 9 MPa·m$^{0.5}$ [30]) the as-cast TiVNbTa exhibits higher fracture toughness even at the lowest temperature tested (~43 MPa·m$^{0.5}$ at -139°C). Other BCC RHEAs with comparable ductility is the HfNbTi(Ta)(Zr) alloy system, which shows good cold-workability and large room temperature tensile strain [12,31–33]. This is due to their heterogeneous grain structure, non-uniform plastic strain distribution, and strain-induced phase transformations (in HfNbTaTiZr). However, TiVNbTa remains single-phase BCC after testing and has large equiaxed grains, suggesting its tensile ductility is likely due to the intrinsic nature of dislocation motion. In addition, Hf-containing alloys have significant restrictions for use in nuclear environments due to the extremely large neutron capture cross-section of Hf. This study provides a preliminary understanding of the thermally-activated plastic deformation of a complex alloy system through activation energy.

In conclusion, the fracture properties of the as-cast single-phase BCC RHEA, TiVNbTa, were investigated using notched 4PB tests. The as-cast TiVNbTa exhibits a soft brittle-to-ductile transition, and the BDTT was found to be between -47°C to -27°C under a constant strain-rate of $10^{-3}$ s$^{-1}$. The activation energy was estimated to be around 0.522±0.095 eV, using an empirical relationship obtained from other BCC refractory metals. Fracture toughness of the as-cast TiVNbTa at -139°C was around 43 MPa·m$^{0.5}$, which is higher than most RHEAs. Despite the brittle fracture surfaces at -139°C, small-scale plasticity was still observed in the load-displacement curves, suggesting some dislocations remained active even at the lowest temperature. Tests within the semi-brittle range exhibited both stable crack growth and plastic deformation, which were supported by observations of the fracture surfaces, showing a combination of ductile (micro-voids, tearing ridges) and brittle (cleavage-like) features. Tests above the BDTT yielded, suggesting dislocations were mobile enough to shield the crack tip effectively, hence reducing the critical stress level required for crack growth. This work demonstrated that the BCC RHEA TiVNbTa in its as-cast state is a potential ductile alloy for structural applications and undoubtedly further processing could enhance its mechanical properties.

**Author Contributions:** *CRediT Contributor Roles Taxonomy*
- R J Scales – Formal Analysis, Investigation, Methodology, Software, Visualisation, Writing – Original Draft
- B-S Li – Supervision, Investigation, Methodology, Visualisation, Writing – Original Draft
- D E J Armstrong – Conceptualisation, Resources, Supervision, Writing – Review & Editing
- A J Wilkinson – Conceptualisation, Supervision, Writing – Review & Editing

**Acknowledgements**

*RJS*, *DEJA*, *AJW* and *BSL* Acknowledge funding from EPSRC grants EP/R021775/1, EP/R006245/1, EP/P001645/1. The authors acknowledge the use of characterisation facilities within the David Cockayne Centre for Electron Microscopy, Department of Materials, University of Oxford, alongside financial support provided by the Henry Royce Institute (Grant ref EP/R010145/1)